\PassOptionsToPackage{table}{xcolor}
\documentclass[sigconf,draft=false]{acmart}
\pdfoutput=1

\makeatletter
\def\@ACM@checkaffil{%
    \if@ACM@instpresent\else
    \ClassWarningNoLine{\@classname}{No institution present for an affiliation}%
    \fi
    \if@ACM@citypresent\else
    \ClassWarningNoLine{\@classname}{No city present for an affiliation}%
    \fi
    \if@ACM@countrypresent\else
        \ClassWarningNoLine{\@classname}{No country present for an affiliation}%
    \fi
}
\makeatother

\ccsdesc[500]{Computing methodologies~Point-based models}
\ccsdesc[300]{Computing methodologies~3D imaging}
\ccsdesc[300]{Computing methodologies~Active vision}
\ccsdesc[500]{Computing methodologies~Uncertainty quantification}

\keywords{neural surface reconstruction, uncertainty quantification}

\setcitestyle{square}

\author{Silvia Sellán}
\affiliation{%
  \institution{University of Toronto}
  \streetaddress{40 St George Street}
  \postcode{M5S 2E4}
  \country{Canada}
}
\email{sgsellan@cs.toronto.edu}
\author{Alec Jacobson}
\affiliation{%
  \institution{University of Toronto \& Adobe Research}
  \country{Canada}
}
\email{jacobson@cs.toronto.edu}

\renewcommand{\shortauthors}{Silvia Sellán and Alec Jacobson}

\usepackage[mathletters]{ucs}
\usepackage[utf8x]{inputenc}

\newcommand{\refequ}[1]{Eq.~(\ref{equ:#1})}

\newcommand{\reffig}[1]{Figure~\ref{fig:#1}}
\newcommand{\refsec}[1]{Section~\ref{sec:#1}}

\newcommand{\new}[1]{{\textcolor{black}{#1}}}
\newcommand{\newmore}[1]{{\textcolor{black}{#1}}}

\providecommand{\D}{}

\providecommand{\I}{}

\providecommand{\K}{}
\providecommand{\L}{}

\providecommand{\N}{}

\providecommand{\P}{}

\providecommand{\R}{}
\providecommand{\S}{}

\renewcommand{\R}{\mathbb{R}}
\renewcommand{\P}{\mathcal{P}}
\renewcommand{\S}{\mathcal{S}}
\renewcommand{\L}{\mathcal{L}}
\renewcommand{\K}{\mathbf{K}}
\renewcommand{\I}{\mathbf{I}}
\renewcommand{\N}{\mathbf{N}}
\renewcommand{\D}{\mathbf{D}}
\newcommand{\grad}{\nabla}

\DeclareMathOperator*{\argmin}{\arg\!\min}
\DeclareMathOperator*{\dx}{dx}
\DeclareMathOperator*{\dtau}{d\tau}
\DeclareMathOperator*{\dxone}{dx_1}
\DeclareMathOperator*{\dxtwo}{dx_2}

\usepackage{tabularx}
\usepackage{booktabs}
\usepackage{makecell}
\usepackage[table]{xcolor}
\definecolor{white}{rgb}{1,1,1}
\definecolor{lightbluishgrey}{rgb}{0.76471,0.84824,0.91647}

\usepackage{subcaption}
\usepackage{wrapfig}

\usepackage{listings}
\usepackage{layouts}
\makeatletter 
\newcommand{\layoutdetails}{%
\begin{tabular}{ll}
 \texttt{\textbackslash{textwidth}} & \printinunitsof{in}\prntlen{\textwidth} \\
\texttt{\textbackslash{linewidth}} & \printinunitsof{in}\prntlen{\linewidth} \\
Main text font &  \f@size pt \f@family \\
\sffamily \small Caption text font &  \sffamily \small \f@size pt \f@family \\
\end{tabular}%
}
\makeatother

\copyrightyear{2023} 
\acmYear{2023} 
\setcopyright{acmlicensed}\acmConference[SA Conference Papers '23]{SIGGRAPH Asia 2023 Conference Papers}{December 12--15, 2023}{Sydney, NSW, Australia}
\acmBooktitle{SIGGRAPH Asia 2023 Conference Papers (SA Conference Papers '23), December 12--15, 2023, Sydney, NSW, Australia}
\acmPrice{15.00}
\acmDOI{10.1145/3610548.3618162}
\acmISBN{979-8-4007-0315-7/23/12}

\begin{document}

\title{Neural Stochastic Screened Poisson Reconstruction}

\begin{abstract}
    Reconstructing a surface from a point cloud is an underdetermined problem. We use a neural network to study and quantify this reconstruction uncertainty under a Poisson smoothness prior. Our algorithm addresses the main limitations of existing work and can be fully integrated into the 3D scanning pipeline, from obtaining an initial reconstruction to deciding on the next best sensor position and updating the reconstruction upon capturing more data.
\end{abstract}

\begin{teaserfigure}
    \includegraphics{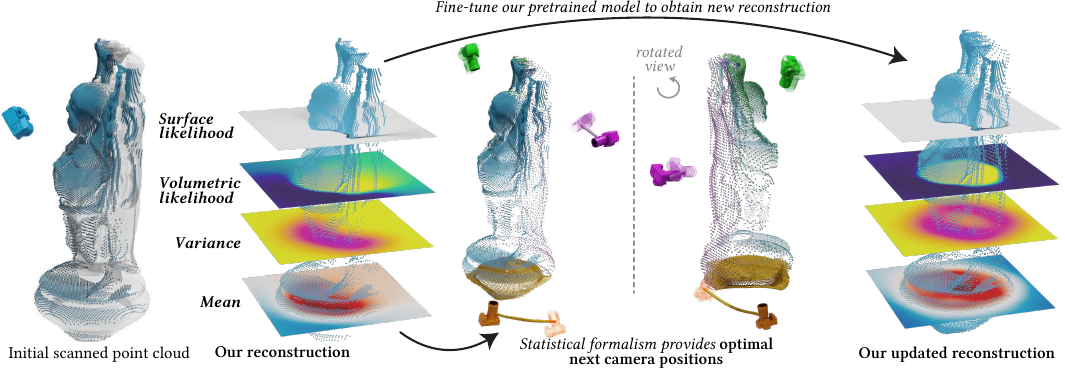}
    \caption{We use a neural network to quantify the reconstruction uncertainty in Poisson Surface Reconstruction (center left), allowing us to efficiently select next sensor positions (center right) and update the reconstruction upon capturing data (right).}\label{fig:teaser}
\end{teaserfigure}

\maketitle

\section{Introduction}

\begin{figure}[b]
  \centering
  \includegraphics{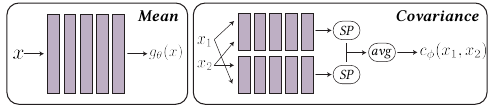}
  \caption{We use neural networks to parametrize the stochastic implicit function describing the reconstructed surface. The mean is a simple five-layered MLP while the covariance includes a SoftPlus (SP) pass and an averaging step to enforce positiveness and symmetry, respectively.}\label{fig:architecture}
\end{figure}

\begin{figure*}
  \centering
  \includegraphics{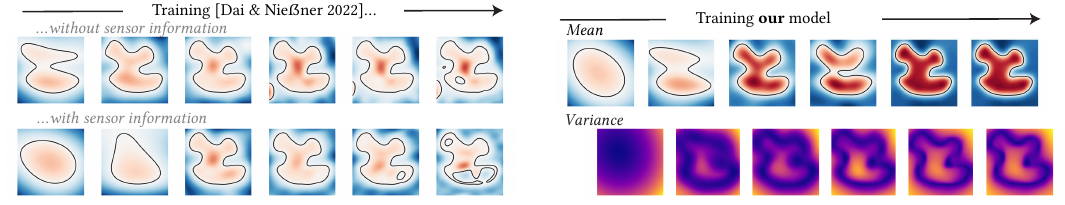}
  \caption{\new{Unlike the Poisson-inspired model by \citet{dai2022neural}, we propose using a neural network to solve the Poisson equation in Poisson Surface Reconstruction, avoiding overfitting in sparsely sampled point clouds. Additionally, we provide a full statistical formalism, including variances (bottom right).}}\label{fig:dai}
\end{figure*}

\emph{Surface reconstruction} is the process of transforming a discrete set of points in space (a common format for captured 3D geometry) into a complete two-dimensional manifold for use in downstream scientific applications. 
Given the fundamentally underdetermined nature of the problem, algorithms must rely on priors to decide on an output surface. 

Absent task-specific knowledge, the predominant geometry processing algorithm for surface reconstruction is \textit{Poisson Surface Reconstruction} (PSR) \cite{kazhdan2006poisson}. PSR encourages smoothness in the reconstruction through a Partial Differential Equation (PDE) whose solution can be computed efficiently and robustly. 
\new{Drawing inspiration from it, \citet{dai2022neural} recently introduced a neural approximation of PSR, which sidesteps the PDE perspective on the problem, achieving some performance gains at the cost of losing theoretical guarantees (see \reffig{dai-vector-field}), overfitting (see \reffig{dai}) and additional  requirements (e.g., sensor positioning).}

Statistically, PSR generates the most probable reconstruction based on the selected prior. This choice inherently defines an entire \emph{posterior} distribution in the space of possible reconstructions. While \emph{Stochastic} PSR \cite{sellan2022stochastic} computes this distribution for the first time in the context of PSR, it demands a complex discretization scheme and relies on multiple approximations to achieve computational tractability.

We build on the work by \citet{sellan2022stochastic} and introduce a neural formulation of Stochastic PSR that provides a full statistical formalism of the reconstruction process while avoiding overfitting and requiring no additional sensor information. Unlike \citet{sellan2022stochastic}, we parametrize the mean and covariance of the implicit field describing the reconstructed surface using a neural network (see \reffig{architecture}), which we optimize using gradient-based optimization on losses derived from the variational version of the Poisson equation. Our neural formulation also allows us to extend this stochastic perspective beyond the original PSR and into \emph{Screened} PSR \cite{kazhdan2013screened}.

We showcase the power of our algorithm by showing its performance in a breadth of applications made possible by our novel neural perspective. In particular, we show how one can fully integrate our algorithm in the 3D scanning pipeline, from obtaining an initial reconstruction 
\new{to defining a differential camera score that can guide the choice of the next best scanning position} and efficiently updating the previous reconstruction (see \reffig{teaser}) \new{by fine-tuning our network with additional data}. We also explore promising avenues for future work, like latent space generalization over scanning positions for a given object or over a space of objects.

\section{Related Work}

\subsection{Surface reconstruction}

Three-dimensional geometry is often captured by recording the distance from a
sensor or \emph{depth camera} to a real-world object
\cite{raj2020survey,ozyecsil2017survey}. Combining the information from many
sensors allows us to represent the raw captured geometry as a discrete set of
points in space or \emph{point cloud}. It is often possible to use properties
about the sensor positioning or heuristics based on global or local cloud
attributes
\cite{hoppe1992surface,konig2009consistent,schertler2017towards,metzer2021orienting}
to equip every point with a normal direction, allow for the slightly more
complete representation of an \emph{oriented point cloud}.

Despite their ubiquitousness, (oriented) point clouds are a fundamentally
underdetermined surface representation: by specifying only a discrete set of
points in space through which a surface passes, it describes a theoretically
infinite number of possible surfaces. \emph{Surface Reconstruction} algorithms
(see \cite{berger2017survey} for a survey) use a \emph{prior} to decide between
them and output a fully determined surface, usually in a format appropriate for
specific downstream tasks like a mesh or an implicit function. These priors
range from simple geometric primitives \cite{schnabel2009completion} to global
properties like symmetry \cite{pauly2008discovering} or self-similarity
\cite{williams2019deep}, user-specified ones  \cite{sharf2007interactive} and,
especially in recent years, data-driven
\cite{remil2017surface,groueix2018papier}.

Absent task-specific knowledge, a commonly used prior is smoothness. This can be
enforced explicitly by considering only surfaces parametrized by a smooth family
of functions; for example, spatially-varying polynomials
\cite{levin2004mesh,alexa2003computing,ohtake2005multi} and linear combinations
of radial basis functions \cite{carr2001reconstruction}. Smoothness can also be
enforced variationally: \emph{Poisson Surface Reconstruction} (PSR)
\cite{kazhdan2006poisson,kazhdan2013screened} encodes \new{volumetric} smoothness \new{away from the input point cloud} by minimizing
the \new{integrated} gradient of the surface's implicit representation and remains one of the
best performing general surface reconstruction algorithms in terms of robustness
and efficiency (see Table 1 in \cite{berger2017survey}). While the authors solve
this optimization problem using the Finite Element Method on a hierarchical
grid, \citet{dai2022neural} have recently proposed using a neural network for a
similar task\new{, albeit they suggest forgoeing the volumetric integration and instead minimizing the gradient only at the point cloud points (see Figs.~\ref{fig:dai} and \ref{fig:dai-vector-field}).}
We cover PSR \new{and its variants} in more detail in \refsec{psr}.

\subsection{Stochastic Surface Reconstruction}

From a statistical perspective, the vast majority of surface reconstruction
works limit themselves to outputting the likeliest surface given the point cloud
observations and their assumed prior. Relatively fewer works take this
stochastic perspective one step further and compute a \emph{posterior}
distribution of all possible surfaces conditioned on the observations. For
example, \citet{pauly2004uncertainty} quantify the likelihood of any spatial
point belonging to the reconstructed surface by measuring its alignment with the
point cloud. 

More recently, \emph{Stochastic Poisson Surface Reconstruction}
\cite{sellan2022stochastic} reinterprets the classic algorithm as a Gaussian
Process, enabling the computation of statistical queries crucial to the
reconstruction process and applications such as ray casting, point cloud repair,
and collision detection. The authors utilize a Finite Element discretization to
compute the mean and covariance functions of the posterior multivariate Gaussian
distribution, which represents the likelihood of all possible reconstructions
(see \refsec{stochastic-psr}), 
\new{resorting to a several approximations and parameter choices for
computational tractability (see \reffig{spsr-comparison}).} In contrast, our work proposes the parametrization
of these functions using neural networks, optimizing them through gradient-based
methods for a more efficient and flexible approach \newmore{while still computing the same statistical quantities (see \reffig{slices}).}

\subsection{Neural PDE solvers}

We propose solving the Poisson equation in Stochastic PSR using a neural
network. As such, our algorithm is one more application in the growing field of
neural partial differential equation solvers. A broad class of these are
\emph{Physics-Informed Neural Networks} (PINNs) (see \cite{cuomo2022scientific}
for a literature review), which effectively soften a PDE and its boundary
conditions into integral loss terms that are minimized with, e.g., stochastic
gradient descent.

If a given PDE accepts a variational formulation, the above process can be done
in a more principled way, as shown by \citet{yu2018deep}. This is the case for
the Poisson equation, which can be equivalently described as a variational
Dirichlet energy minimization. This is noted by \citet{sitzmann2019siren}, who
show the impressive performance of sinusoidal activation functions when applied
to Dirichlet-type problems. We borrow from their observations and propose a
network architecture with sine activations.

\begin{figure}
    \centering
    \includegraphics{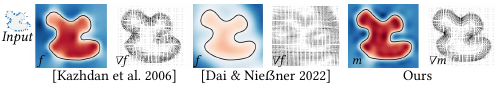}
    \caption{Even before overfitting, the result by \citet{dai2022neural} does not replicate the PSR output, with non-zero gradients away from the data.}\label{fig:dai-vector-field}
  \end{figure}

\subsection{Next-Best-View planning}

A key benefit of proposed approach is its integration in the 3D scanning
process. Specifically, it allows us to compute a \emph{score} function that
quantifies how useful a proposed next sensor position would be for the
reconstruction task. This is a common first step in the \emph{active vision} or
\emph{next-best-view planning} pipeline, which has been a subject of study for
decades (see, e.g., \cite{scott2003view,chen2011active} for surveys). In it,
prospective sensor placements may be scored by accounting for one or several
factors like coverage
\cite{connolly1985determination,yamauchi1997frontier,bircher2016receding},
navigation distance, expected reconstruction error \cite{vasquez2014volumetric}, \new{scene segmentation entropy \cite{xu2015autoscanning}}
and redundancy of multiple views \cite{Lauri_RAL2020}. Orthogonally, works may
need to rely on coarse shape priors for the reconstruction
\cite{zhou2020offsite,zhang2021continuous} or balance improving reconstruction
in sampled areas with exploring new unsampled ones.

More recently, volumetric methods like those by \citet{isler2016information},
and \citet{daudelin2017adaptable} use simple heuristics (e.g., distance to the
point cloud combined with visibility) to quantify the marginal likelihood of a
given point in space being contained in the reconstructed object. This quantity
is discretized onto a voxel grid and used to quantify the expected information
gain from a given sensor position. Building on these works, our proposed utility
function requires no heuristics, coming instead directly from the statistically
formalized reconstruction process and, unlike \cite{daudelin2017adaptable},
accounts for the possible spatial interdependencies along a single ray (see
\reffig{ray-casting-didactic}). Further, since our reconstruction is
parametrized by a neural network, this score is differentiable with respect to
the sensor parameters, allowing for the efficient discovery of locally optimal
camera placements (see \reffig{camera-backwards}). While we introduce said novel,
differentiable utility function, the development of a comprehensive
next-best-view planning pipeline, which would encompass global searches, travel
times, collision avoidance, and robot constraints, falls outside the scope of
this paper.

Finally, outside of the point cloud reconstruction realm, the recent popularity
of Neural Radiance Fields \cite{mildenhall2021nerf} has also given rise
to uncertainty-driven approaches for next-best-view planning in RGB multi-view
representations (see, e.g.,
\cite{smith2022uncertainty,sucar2021imap,kong2023vmap,jin2023neu}).

\begin{figure}
    \centering
    \includegraphics{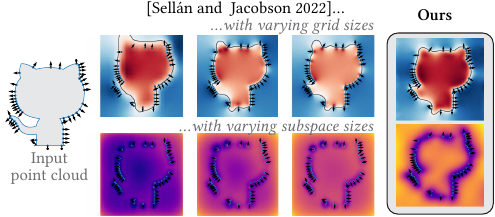}
    \caption{\new{\citet{sellan2022stochastic} couple the reconstruction lengthscale with their discretization grid spacing, and require a subspace approximation. Our neural network discretizations avoids both issues.}}\label{fig:spsr-comparison}
\end{figure}

\begin{figure*}
    \centering
    \includegraphics{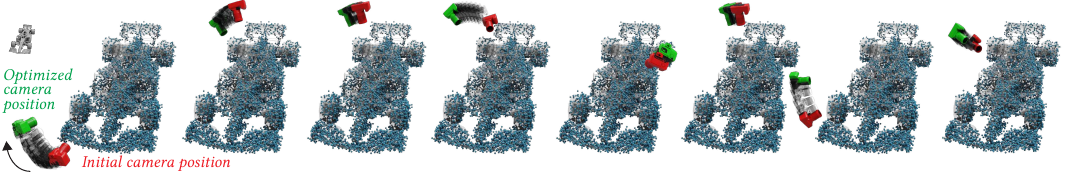}
    \caption{\new{We provide a differentiable utility function that we can optimize to explore local next-best-views.}}\label{fig:camera-backwards}
\end{figure*}

\section{Background}

Given an oriented point cloud $\P$ with points $p_1,\dots,p_n$ and
corresponding (outward-facing) normal observations $\vec{n}_1,\dots,\vec{n}_n$, we consider the
implicit reconstruction task of finding a function $f:\R^d\rightarrow \R$ such
that
\begin{equation}\label{equ:reconstruction}
  f(p_i) = 0\,,\qquad \grad f(p_i) = \vec{n}_i\,,\qquad \forall
  i\in\{1,\dots,n\}\,.
\end{equation}
The zero levelset $\S=f^{-1}(\{0\})$ is the reconstructed surface, whose interior is $\Omega =\{x\in\R^d\,:\,f(x)\leq 0\}$.
\new{We will be consistent with this convention that places \emph{negative} implicit function values \emph{inside} the reconstruction, and \emph{positive} ones \emph{outside}.}

\begin{figure}
  \centering
  \includegraphics{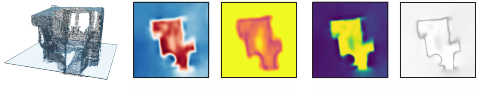}
  \includegraphics{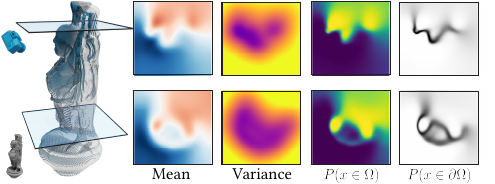}
  \caption{\new{Like \citet{sellan2022stochastic}, our algorithm can respond to statistical queries related to the reconstruction.}}\label{fig:slices}
\end{figure}

\subsection{Poisson Surface Reconstruction}\label{sec:psr}

\emph{Poisson Surface Reconstruction} (PSR) \cite{kazhdan2006poisson} builds $f$ in
two steps. First, a smear kernel $F$ is used to interpolate $\vec{n}_i$ into a
vector field $\vec{v}:\R^d\rightarrow \R^d$ defined in a box $B$ containing $\P$:
\begin{equation}
  \vec{v}(x) = \sum\limits_{i=1}^{n}F(x,x_i)\vec{n}_i\,.
\end{equation}
Then, $f$ is defined as the function whose gradient best matches
$\vec{v}$:
\begin{equation}\label{equ:dirichlet}
  f = \argmin_{g}\int_B \|\vec{v}(x) - \grad g(x)\|^2\,\dx\,.
\end{equation}
This variational problem is equivalent to the Poisson equation
\begin{equation}\label{equ:poisson}
  \Delta f = \grad \cdot \vec{v}(x)\,,
\end{equation}
which the authors discretize using the Finite Element Method on an octree and
solve using a purpose-built multigrid algorithm. Since \refequ{poisson} alone
does not uniquely determine $f$, a valid $f$ is computed and its values shifted
to best satisfy $f(p_i)=0$.

In \emph{Screened Poisson
Surface Reconstruction}, \citet{kazhdan2013screened} circunvent this by adding a
\emph{screening} term to \refequ{dirichlet}

\begin{equation}
  f = \argmin_{g}\int_B \|\vec{v}(x) - \grad g(x)\|^2\,\dx \,+\, \lambda \sum\limits_{i=1}^n
  g(p_i)^2
\end{equation}
which translates into a Screened Poisson equation
\begin{equation}
  (\Delta - \lambda I) f = \grad \cdot \vec{v}(x)\,,
\end{equation}
for a specific masking operator $I$.

\subsection{Stochastic Poisson Surface Reconstruction}\label{sec:stochastic-psr}

Screened or not, the output of Poisson reconstruction is a single function $f$.
However, the reconstruction task is fundamentally uncertain: 
\new{\refequ{reconstruction} alone is underdetermined and satisfied by an infinite number of possible functions $f$. When subject to appropriate boundary conditions, Poisson reconstruction selects one particular solution, which can be understood as the most likely solution under a given prior. } \citet{sellan2022stochastic} formalize this
statistical intuition by interpreting $(p_i,\vec{n}_i)$ as observations of a Gaussian
Process and computing the posterior distribution

\begin{equation}
  \vec{v}\,|\,(p_1,\vec{n}_1),\dots,(p_1,\vec{n}_n)\sim
  \mathcal{N}(\vec{\mu}(x),\Sigma(x,x'))\,.
\end{equation}

\refequ{dirichlet} is then enforced in the space of distributions, obtaining a
posterior for $f$,

\begin{equation}
  f\,|\,(p_1,\vec{n}_1),\dots,(p_1,\vec{n}_n)\sim
  \mathcal{N}(m(x),k(x,x'))\,,
\end{equation}

whose mean and covariance functions $m,k$ are solutions to the variational
problem

\begin{align}
  m & = \argmin_{g}\int_B \|\vec{\mu}(x) - \grad g(x)\|^2\,\dx\,,
  \label{equ:mean-variational}\\
  k & = \argmin_{c}\iint_B \|\Sigma(x_1,x_2) - \mathbf{D} c(x_1,x_2)
  \|_F^2\,\dxone\dxtwo\,,\label{equ:cov-variational}
\end{align}
where $\mathbf{D} c(x_1,x_2)$ is the $d\times d$ matrix whose $i,j$ entries are
\begin{equation}
  \left.\frac{\partial^2}{\partial a_i \partial b_j} c(a,b)\right|_{a=x_1,b=x_2}\label{equ:d-operator}
\end{equation}

In the same way of \refequ{dirichlet}, Eqs.~(\ref{equ:mean-variational}) and
(\ref{equ:cov-variational}) can be written as Poisson-style PDEs that are solved using the Finite Element Method
on a uniform or hierarchical grid. Like the original work by
\citet{kazhdan2006poisson}, \citet{sellan2022stochastic} shift the values of $m$ and $k$ after
the fact to satisfy $m(p_i)=k(p_i,p_i)=0$ on average.

\section{Method}

We propose discretizing $g$ and $c$ in Eqs.~(\ref{equ:mean-variational}) and
(\ref{equ:cov-variational}) using neural networks parametrized by weights
$\theta$ and $\phi$ and solving them directly using gradient-based optimization.

\subsection{Loss}

\begin{figure}
  \centering
  \includegraphics{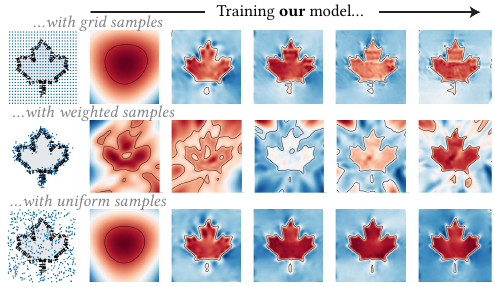}
  \caption{\new{We choose to draw samples uniformly from a bounding box around the point cloud after observing overfitting when using different strategies.}}\label{fig:sampling}
\end{figure}

Given $s$ samples $x_1,\dots,x_s\in\R^d$ drawn from a uniform distribution of $B$ \newmore{(see \reffig{sampling})}, let us define the \emph{Dirichlet mean loss} as
\begin{equation}\label{equ:dirichlet-mean-loss}
  \L^m_{D}(\theta) = \frac{|B|}{s} \sum_{i=1}^s\|\vec{\mu}(x_i) - \grad g_\theta(x_i)\|^2
\end{equation}
and its covariance counterpart
\begin{equation}\label{equ:dirichlet-covariance-loss}
  \L^k_{D}(\phi) = \frac{|B|}{s} \sum_{i=1}^s\sum_{j=1}^s\|\Sigma(x_i,x_j) - \mathbf{D} c_\phi(x_i,x_j)\|_F^2\,.
\end{equation}
By Monte Carlo integration, we have
\begin{equation}\label{equ:mc1}
  \L^m_{D}(\theta) \approx \int_B \|\vec{\mu}(x) - \grad g_\theta(x)\|^2\,\dx
\end{equation}
and
\begin{equation}\label{equ:mc2}
  \L^k_{D}(\phi) \approx \iint_B \|\Sigma(x_1,x_2) - \mathbf{D} c_\phi(x_1,x_2)\|_F^2 
  \,\dxone\dxtwo\,.
\end{equation}
Thus, the functions $g_{\theta^\star}$ and $c_{\phi^\star}$ parametrized by the minimizers
\begin{equation}
  \{\theta^\star,\phi^\star\} = \argmin_{\theta,\phi} \L^m_{D}(\theta) + \L^k_{D}(\phi) 
\end{equation}
are solutions to the variational problem in  Eqs.~(\ref{equ:mean-variational}) and (\ref{equ:cov-variational}) when restricted to the space of neural-network-parametrized functions. Thus, they are also Poisson solutions.

It should be noted that, if one substitutes the samples $x_i$ with the points in the input point cloud $p_i$ in \refequ{dirichlet-mean-loss}, $\mathcal{L}_D^m(\theta)$ is identical to the loss proposed by \citet{dai2022neural}. However, our decoupling of the sampling from the point cloud is critical.
\new{Importantly, it is only by sampling from the volumetric bounding box in \refequ{dirichlet-mean-loss} that we can claim to be approximating the volumetric integral in \refequ{mc1} and thus solving a Poisson equation. Theoretically, this choice has the effect of making our algorithm into a strict generalization of PSR (see \reffig{dai-vector-field}); in practice, it imposes a volumetric smoothness prior that avoids overfitting (see \reffig{dai}).}

\begin{figure}
  \centering
  \includegraphics{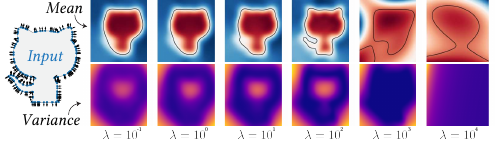}
  \caption{Our screen weight balances smoothness with input fidelity, but can complicate convergence at high values.}\label{fig:ablation-lambda}
\end{figure}

An immediate benefit of this neural perspective is the possibility to extend the statistical formalism of \citet{sellan2022stochastic} from the original Poisson Surface Reconstruction \cite{kazhdan2006poisson} to its improved, screened version \cite{kazhdan2013screened}. We can do so merely by adding mean and covariance \emph{screen losses}
\begin{equation}\label{equ:screen-loss}
  \mathcal{L}^m_S(\theta) = \frac{1}{n}\sum_{i=1}^n\|g_\theta(p_i)\|^2\,,\quad \mathcal{L}^k_S(\phi) = \frac{1}{n}\sum_{i=1}^n\|c_\phi(p_i,p_i)\|^2\,,
\end{equation}
which we combine with the Dirichlet losses to reach our total loss
\begin{equation}\label{equ:total-loss}
  \mathcal{L}(\theta,\phi) =  \L^m_{D}(\theta) + \L^k_{D}(\phi) + \lambda_S \mathcal{L}^m_S(\theta) + \lambda_S \mathcal{L}^k_S(\phi)
\end{equation}
Inspired by the choice made by \citet{dai2022neural}, which we validate experimentally (see \reffig{ablation-lambda}), we fix $\lambda_S=100$.

\begin{figure}
  \centering
  \includegraphics{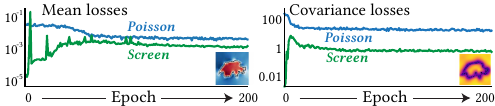}
  \caption{For our choice of hyperparameters, the Poisson losses regularly dominate over the screening terms.}\label{fig:losses}
\end{figure}

\subsection{Data generation}\label{sec:sampling}

To evaluate $\mathcal{L}(\theta,\phi)$, we first choose $B$ to be a loose box around the input point cloud and \new{uniformly} sample $x_1,\dots,x_s\in B$. Then, as described by \citet{sellan2022stochastic}, we compute the matrices
\begin{equation}
  \K_1 = (F(x_i,x_j))_{i,j}\in\R^{s\times s}\,,\quad \K_2 = (F(x_i,p_j))_{i,j}\in\R^{s\times n}\,,
\end{equation}
as well as the lumped sample covariance matrix
\begin{equation}
  \D \approx \K_3 = (F(p_i,p_j))_{i,j}\in\R^{n\times n}\, .
\end{equation}
We employ the same approximated Gaussian kernel suggested by the authors and make use of its compact support to efficiently evaluate the above matrices with a KD tree.
Using these matrices, we compute the Gaussian Process posterior mean
\begin{equation}
  \pmb{\mu} = \K_2 \D^{-1} \N\,,
\end{equation}
where $\N\in\mathbb{R}^{n\times d}$ concatenates $\vec{n}_1,\dots,\vec{n}_n$, and the covariance
\begin{equation}
  \pmb{\Sigma} = \K_1 - \K_2 \D^{-1} \K_2^\top\,.
\end{equation}

The row entries in $\pmb{\mu}$ then correspond to $\vec{\mu}(x_i)$, while each scalar entry in $\pmb{\Sigma}$ determines the $d\times d$ matrix $\Sigma$ through $\Sigma(x_i,x_j) = \pmb{\Sigma}_{i,j}\I$.
\new{As we validate experimentally in \reffig{sampling}, sampling $B$ uniformly during training is necessary to maintain the theoretical guarantees in Eqs. \ref{equ:mc1} and \ref{equ:mc2}. More elaborate strategies beyond this work's scope (e.g., Metropolis-Hastings integration) that would result in weights being added in Eqs. \ref{equ:mc1} and \ref{equ:mc2} may yield performance improvements.}

\subsection{Architecture \& Training}

\begin{figure}
    \centering
    \includegraphics{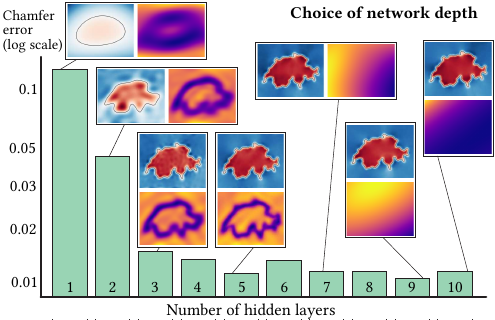}
    \caption{\new{Too few hidden layers can limit the geometric detailed captured un our reconstructions. At the same time, we observe diminishing returns and difficuly with covariance convergence for higher layer numbers.}}\label{fig:network-depth}
  \end{figure}

We model $g_\theta$ and $c_\phi$ using two five-layered MLPs \new{(see \reffig{network-depth})} with 512 internal hidden units and sine activation functions \cite{sitzmann2019siren}. Our covariance network $c_\phi$ also includes a SoftPlus layer to enforce positivity, followed by an averaging $(c_\phi(x_1,x_2) + c_\phi (x_2,x_1))/2$ 
\new{(see \reffig{architecture}). Combined with Schwarz's theorem, this forces  $\mathbf{D} c(x_1,x_2)$ in \refequ{mc2} to be symmetric by construction.} We experimented with residual connection layers as suggested by \citet{yu2018deep}, but found no significant performance improvement.

At each epoch, we generate 100,000 covariance and 100,000 mean Poisson samples $x_i$ together with an equal number of screening samples selected from the point cloud $p_i$ (with repetition if necessary) as detailed in \refsec{sampling}. 
\new{This sampling results in four datasets (covariance, mean, covariance screening and mean screening). We cycle through all four}
with repetition until they are all exhausted with a 512 batch size, evaluating our losses and backpropagating through them to compute the gradient of $\L(\theta,\phi)$ with respect to $(\theta,\phi)$. We then use the Adam \cite{kingma2014adam} optimizer with learning rate $10^{-4}$ and weight decay $10^{-5}$. We repeat this process for a number of epochs between 50 and 200 (see \reffig{losses}).

\paragraph{Implementation details} We implement our algorithm in \textsc{Python}, using \textsc{PyTorch} to build and train our model and \textsc{Gpytoolbox} \cite{gpytoolbox} for common geometry processing subroutines. In our 3.0Ghz 18-core Linux machine with a 48 GB NVIDIA RTX A6000 graphics card and 528 GB RAM, our unoptimized implementation lasts around 30 seconds to train each epoch, the main bottleneck being the backpropagation through the $\mathbf{D}$ operator in \refequ{d-operator}. For Figures~\ref{fig:dai} and \ref{fig:dai-vector-field}, we implemented the algorithm by \citet{dai2022neural} following their instructions in the absence of author-provided code. We rendered our 3D results using \textsc{Blender}.

\section{Results \& Applications}

\subsection{3D Scanning integration}

Once a point cloud has been captured, our method can be used to compute all kinds of statistical queries useful to the reconstruction (\reffig{slices}) in the same way as described by \citet{sellan2022stochastic}. However, our novel neural perspective goes one qualitative step further and allows for a full integration into the scanning process, providing feedback over where to scan next and efficiently updating a given reconstruction upon capturing more data.

\subsubsection{Ray casting} 

Given a captured scan and a proposed sensor position $\mathbf{r}$ and direction $\mathbf{d}$, a crucial question is where a ray travelling from the sensor would intersect the surface. In traditional volumetric rendering terms, this amounts to computing the \emph{opacity} along the ray, or the likelihood that a ray emanating from the sensor reaches a given distance without terminating.

\citet{sellan2022stochastic} suggest computing the marginal probabilities along the ray
\begin{equation}
   p(t) = P(f(\mathbf{r} + t\mathbf{d})\leq 0)
\end{equation} 
and interpreting these as densities
\begin{equation}
    \rho(t) = \frac{p(t)}{1 - p(t)}
\end{equation}
that they propose integrating to compute the opacity
\begin{equation}\label{equ:opacity-bad}
    o(t) = 1 - e^{-\int_0^t \rho(\tau)\dtau}\,.
\end{equation}

However, we note that this expression for the opacity is usually employed in the context of gases, for which the effects of inter-particle interactions are negligible and one can assume that the likelihood of encountering a gas particle at time $\tau$ is independent of encountering one at time $\tau + \dtau$, giving validity to the integral in \refequ{opacity-bad}. This independence assumption does not hold for the case of uncertain solids, as evidenced by \reffig{ray-casting-didactic}: while the marginal likelihood is $p(t)=0.5$ for all $t$ between $t_1$ and $t_2$, there is no configuration of the shape for which a ray terminates at $t$. Statistically, this is because the point at time $t$ is fully correlated with the point at time $t_1$. While \reffig{ray-casting-didactic} is an extreme example, this difference appears in general reconstruction examples (see \reffig{transmittances-didactic})

Accounting for these correlations is simple. Instead of \refequ{opacity-bad}, one can compute the opacity as the joint probability that $f$ was positive at every point in the ray prior to $\mathbf{r} + t\mathbf{d}t$:
\begin{equation}\label{equ:opacity}
    o(t) = P(f(\mathbf{r} + \tau\mathbf{d})>0\,,\forall \tau < t)\,.
\end{equation}
We uniformly discretize the interval $[0,t]$ such that it amounts to querying a cumulative multivariate Gaussian.
\new{Fortunately, as shown by \citet[Sec. 6]{marmin2015differentiating}, this expression can be differentiated with respect to the entries in the covariance matrix with the aid of Plackett's formula \cite{berman1987extension}. We use the \textsc{PyTorch} implementation of this formula by \citet{marmincode} for this task.}

\begin{figure}
    \centering
    \includegraphics{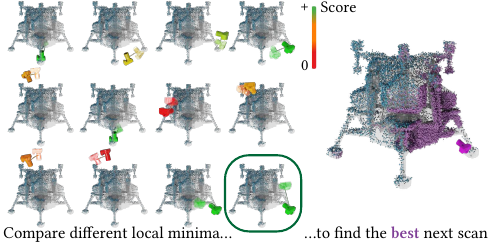}
    \caption{\new{Our local next-best-view search is best combined with a global search, where the scores of different local optima are compared.}}\label{fig:next-view-planning}
\end{figure}

\subsubsection{Next view planning}

As seen above, the time travelled by a ray from a given camera position before colliding with the surface can be interpreted as a random variable, whose cumulative distribution function is the opacity in \refequ{opacity}. Crucially, by Foubini's theorem, this means one can compute the expected collision time as
\begin{equation}
    \left\langle t(\mathbf{r},\mathbf{d}) \right\rangle = \int_0^\infty \left(1 - o(\tau)\right)\dtau\,,
\end{equation}
leading to the expected collision point
\begin{equation}
    \mathbf{p}^\star (\mathbf{r},\mathbf{d}) = \mathbf{r} + \left\langle t(\mathbf{r},\mathbf{d}) \right\rangle \mathbf{d}\,.
\end{equation}

The optimal sensor position will be one that generates a new point cloud point in an area of high variance. Therefore, it makes sense to define the \emph{score} of a camera as
\begin{equation}
    u(\mathbf{r},\mathbf{d}) = \sigma(\mathbf{p}^\star (\mathbf{r},\mathbf{d})) = c_{\phi^\star}(\mathbf{p}^\star (\mathbf{r},\mathbf{d}),\mathbf{p}^\star (\mathbf{r},\mathbf{d}))\,.
\end{equation}
While \citet{sellan2022stochastic} propose a camera scoring criteria, our novel neural perspective allows us to backpropagate through $c_\phi$, meaning that we can compute the gradient of the score with respect to camera parameters $(\mathbf{r},\mathbf{d})$, and find an optimal camera position with gradient descent. We show the potential of this contribution in \reffig{camera-backwards}, inspired by Fig. 26 by \citet{sellan2022stochastic}.

This gradient-based next view angle optimization will often converge to suboptimal local minima. Indeed, as we show in \reffig{next-view-planning}, it is better combined with a global search by sampling several initial sensor positions, backpropagating to find an optimum near them, and then choosing the converged camera with the best global score.
\new{In \reffig{nbv-error-plot}, we quantify the quality of our subsequent chosen views of a mechanical object by showing they improve on randomly sampled ones. Only in this simplified setup in which views are sampled from a sphere around the object and the directions are constrained to aim to the same spatial point, we are able to compare also to other heuristics like furthest-point sampling, which we show our more generally applicable method matches or outperforms.}

\begin{figure}
    \centering
    \includegraphics{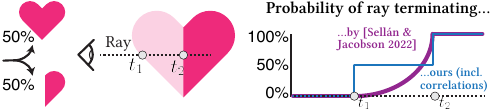}
    \caption{\citet{sellan2022stochastic} consider only marginal likelihoods to compute the termination probability along a ray. This leads to inaccuracies in cases with high correlations among spatial points (see text).}\label{fig:ray-casting-didactic}
  \end{figure}

\begin{figure}
    \centering
    \includegraphics{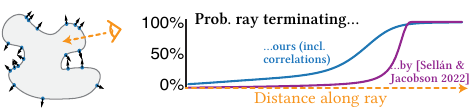}
    \caption{Accounting for correlations leads to significant differences in the ray termination distribution for general point cloud reconstruction examples.}\label{fig:transmittances-didactic}
\end{figure}

\begin{figure}
    \centering
    \includegraphics{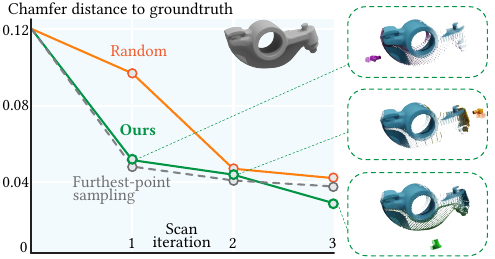}
    \caption{\new{In next-view selection, our algorithm outperforms random sampling and even matches or improves on commonly used heuristics like furthest point in simple setups when the latter are available.}}\label{fig:nbv-error-plot}
\end{figure}

\subsubsection{Fine-tuning}
Once a new sensor position is chosen and a new scan is taken, points are added to the cloud. Traditional algorithms like PSR would then require investing in an updated discretization and entirely new Poisson solve to obtain an updated reconstruction.

\begin{figure}
    \centering
    \includegraphics{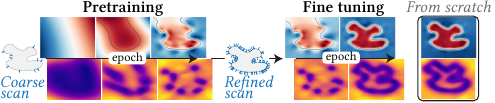}
    \includegraphics{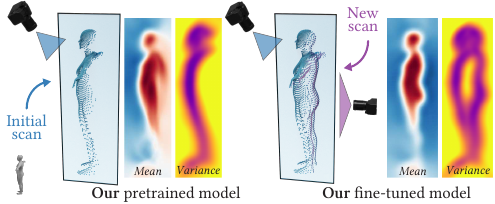}
    \caption{Sequential scanning setups benefit from our model, whose reconstruction can be updated when the object is observed from new angles.}\label{fig:retraining}
  \end{figure}

Fortunately, our neural perspective allows us to take advantage of an earlier reconstruction to update it more efficiently. Indeed, as shown in \reffig{retraining}, we may consider our model's training on the initial point cloud as a \emph{pretraining} of our model, which is \emph{fine-tuned} for only a few epochs every time new points are captured. 

Our model can thus be integrated in an end-to-end scanning pipeline, as once an updated mean and variance is obtained, the best next view angle optimization can start again (see \reffig{teaser}). Our algorithm can even provide a stopping criterion, in the form of the integrated uncertainty proposed by \citet{sellan2022stochastic}.

\subsection{Generalization}

Another major advantage of our neural formalism over a traditional one is the possibility to train our network on many given reconstructions and trust it to generalize to similar-yet-unseen data. This can circumvent expensive optimizations in cases where one has access to a large training set of point clouds and must quickly make inference on a newly observed set of points.

We show a prototypical example of what such a process could look like in \reffig{generalization-2d}, where a training set of point clouds is captured by scanning a shape from several different angles. Our model is then expanded to accept a latent encoding $z$, the values of which are trained simultaneously with the model parameters in the ``autodecoder'' style proposed by \citet{park2019deepsdf}. When a new scan $\mathcal{S}$ of the object is captured, test-time optimization (with the model parameters frozen) produces an optimal latent encoding for the new point cloud. This reconstruction can be used as-is or fine-tuned for a very limited number of epochs for a final reconstruction.

We believe this generalization capability can prove useful in industrial applications, where one may be able to produce a number of partial training scans of an object. Then, objects on an assembly line can be quickly scanned and projected into the learned latent space of partial scans. As we show in \reffig{fandisk}, our model's statistical formalism can then be used (in the form of the point cloud's average log likelihood) to identify foreign objects or defective pieces.

One can also use our model to generalize over a space of different-yet-similar shapes, as we show in \reffig{smpl-generalization}, where a latent space of scans is learned over 20 diverse human scans generated using STAR \cite{osman2020star}. Upon capturing a new scan, test-time latent code optimization can efficiently provide a novel reconstruction.

\section{Limitations \& Conclusion}

As we have shown, a key advantage of our neural formulation is the possibility to iteratively fine-tune reconstructions upon capturing more data. To fully take advantage of our method's efficiency, one may need to optimize its runtime, which we did not do beyond asymptotics. We believe the clearest avenues for speedups are exploring non-uniform distributions for data generation and task-specific weight initializations.

We introduce a method for formalizing reconstruction uncertainty using a neural network. However, it should be noted that this uncertainty is encoded by the Gaussian Process used to generate data, while the network is merely solving a PDE. A promising avenue for future work is circumventing the GP altogether, using Machine Learning uncertainty quantification techniques to obtain a posterior distribution directly from the input point cloud. While this may mean deviating from Poisson Surface Reconstruction, it could present a major improvement in accuracy (removing the need for covariance matrix lumping) and applicability (allowing for sensor-specific non-Gaussian noise patterns).

\begin{figure}
    \centering
    \includegraphics{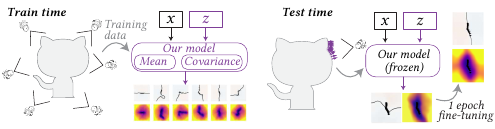}
    \caption{In cases where several scans of an object are available, our model can be combined with autodecoder to efficiently reconstruct new views via test-time optimization.}\label{fig:generalization-2d}
\end{figure}

\begin{figure}
    \centering
    \includegraphics{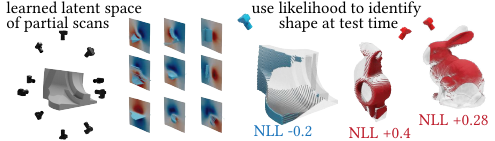}
    \caption{In an industrial setting, one can use our algorithm to learn a latent space of partial scans of an object in order to detect anomalies through any point cloud's negative log likelihood (NLL).}\label{fig:fandisk}
\end{figure}

\begin{figure}
    \centering
    \includegraphics{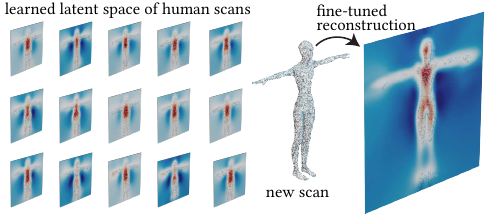}
    \caption{\new{A dataset of similar shapes can be used to learn a latent space of possible scans onto which new scans can be efficiently projected.}}\label{fig:smpl-generalization}
\end{figure}

\newmore{All our generalization results (Figures \ref{fig:generalization-2d}, \ref{fig:fandisk} and \ref{fig:smpl-generalization}) use identical (virtual) scanning devices, and every input point cloud is re-scaled to the unit cube; as such, we do not expect our results to generalize beyond these choices. Future work could mitigate this; for example, by learning a latent space of device parameters and positions as suggested by  \citet{martin2021nerf} in the context of NeRF.}

While uncertainty quantification has become a common consideration in neighboring fields like Computer Vision and Robotics \cite{kendall2017uncertainties}, it remains rare for Computer Graphics works to expose their algorithmic uncertainties. It is our hope that as our tool set grows and our field's application realm diversifies, our work can serve as a first step in the right direction.


\begin{acks}
This project is funded in part by NSERC Discovery (RGPIN2017-05235, RGPAS-2017-507938), New Frontiers of Research Fund (NFRFE-201), the Ontario Early Research Award program, the Canada Research Chairs Program, a Sloan Research Fellowship and the DSI Catalyst Grant program. The first author is funded in part by an NSERC Vanier Scholarship.

We thank Kirill Serkh, Kiriakos Kutulakos, David Lindell, Eitan Grinspun, David I.W. Levin, Oded Stein, Andrea Tagliasacchi, Otman Benchekroun, Lily Goli and Claas A. Voelcker for insightful conversations that inspired us in this work; Hsueh-Ti Derek Liu for his help rendering our results; as well as Rafael Rodrigues (Fig. 6, CC BY-NC-SA 4.0) and  ShaggyDude (Fig. 13, CC BY 4.0) for releasing their 3D models for academic use. We would also like to thank Xuan Dam, John Hancock and all the University of Toronto Department of Computer Science research, administrative and maintenance staff.

\end{acks}

\bibliographystyle{ACM-Reference-Format}
\bibliography{references.bib}

\end{document}